\RequirePackage{rotating}
\documentclass[fleqn,usenatbib]{mnras}
\usepackage{epsfig}
\usepackage{amsmath}
\usepackage{graphicx}
\usepackage{array}
\usepackage{textcomp}
\usepackage{amssymb}
\usepackage{rotating}
\usepackage{pdflscape}
\usepackage{caption}
\usepackage{subcaption}
\usepackage{longtable}
\usepackage{xtab}
\usepackage{xcolor}
\usepackage{colortbl}
\usepackage{hhline}
\def\apjs {ApJS}
\def\apjl {ApJL}

\def\aap {A\&A}

\title[Stellar kinematics of main sequence galaxies]
      {A SAMI and MaNGA view on the stellar kinematics of galaxies on the star-forming main sequence}  
\author[A.\ Fraser-McKelvie et al.]
       {A.\ Fraser-McKelvie$^{1,2}$\thanks{a.fraser-mckelvie@uwa.edu.au}, L.\ Cortese$^{1,2}$, J.\ van de Sande$^{2,3}$, J.\ J.\ Bryant$^{2,3,4}$, B.\ Catinella$^{1,2}$,  M.\ Colless$^{2,5}$, \and S.\ M.\ Croom$^{2,3}$, B.\ Groves$^{1,2}$, A.\ M.\ Medling$^{6,2}$, N.\ Scott$^{2,3}$, S.\ M.\ Sweet$^{2,7}$, 
       J.\ Bland-Hawthorn$^{3}$, \and M.\ Goodwin$^{8}$,  J.\ Lawrence$^{8}$,  N.\ Lorente$^{9}$,  M.\ S.\ Owers$^{10,11}$, S.\ N.\ Richards$^{12}$.        
                \vspace*{1mm}\\
        $^{1}$ International Centre for Radio Astronomy Research, The University of Western Australia, 35 Stirling Hwy, 6009 Crawley, WA, Australia \\
        $^{2}$ ARC Centre of Excellence for All Sky Astrophysics in 3 Dimensions (ASTRO 3D) \\
        $^{3}$ Sydney Institute for Astronomy (SIfA), School of Physics, The University of Sydney, NSW 2006, Australia \\
        $^{4}$ Australian Astronomical Optics, AAO-USydney, School of Physics, University of Sydney, NSW 2006, Australia \\
        $^{5}$ Research School of Astronomy and Astrophysics, Australian National University, Canberra, ACT 2611 \\
        $^{6}$ Ritter Astrophysical Research Center, University of Toledo, Toledo, OH 43606, USA \\
        $^{7}$ School of Mathematics and Physics, University of Queensland, Brisbane, QLD 4072, Australia \\
        $^{8}$ Australian Astronomical Optics -- Macquarie, 105 Delhi Rd, North Ryde, NSW 2113, Australia \\
        $^{9}$ AAO-MQ, Faculty of Science \& Engineering, Macquarie University. 105 Delhi Rd, North Ryde, NSW 2113, Australia \\
        $^{10}$ Department of Physics and Astronomy, Macquarie University, NSW 2109, Australia \\
        $^{11}$ Astronomy, Astrophysics and Astrophotonics Research Centre, Macquarie University, Sydney, NSW 2109, Australia \\
        $^{12}$ SOFIA Science Center, USRA, NASA Ames Research Center, Building N232, M/S 232-12, P.O. Box 1, Moffett Field, CA 94035-0001, USA \\
	}
\begin{document}
\maketitle
\begin{abstract}
Galaxy internal structure growth has long been accused of inhibiting star formation in disc galaxies. We investigate the potential physical connection between the growth of dispersion-supported stellar structures (e.g. classical bulges) and the position of galaxies on the star-forming main sequence at $z\sim0$. Combining the might of the SAMI and MaNGA galaxy surveys, we measure the $\lambda_{Re}$ spin parameter for 3289 galaxies over $9.5 < \log M_{\star} [\rm{M}_{\odot}] < 12$. At all stellar masses, galaxies at the locus of the main sequence possess $\lambda_{Re}$ values indicative of intrinsically flattened discs. 
However, above $\log M_{\star}[\rm{M}_{\odot}]\sim10.5$ where the main sequence starts bending, we find tantalising evidence for an increase in the number of galaxies with dispersion-supported structures, perhaps suggesting a connection between bulges and the bending of the main sequence. 
Moving above the main sequence, we see no evidence of any change in the typical spin parameter in galaxies once gravitationally-interacting systems are excluded from the sample.  
Similarly, up to 1 dex below the main sequence, $\lambda_{Re}$ remains roughly constant and only at very high stellar masses ($\log M_{\star}[\rm{M}_{\odot}]>11$), do we see a rapid decrease in $\lambda_{Re}$ once galaxies decline in star formation activity.
If this trend is confirmed, it would be indicative of different quenching mechanisms acting on high- and low-mass galaxies. The results suggest that while a population of galaxies possessing some dispersion-supported structure is already present on the star-forming main sequence, further growth would be required after the galaxy has quenched to match the kinematic properties observed in passive galaxies at $z\sim0$.

\end{abstract}

\begin{keywords}
 galaxies: evolution -- galaxies: general  -- galaxies: bulges -- galaxies: kinematics and dynamics
\end{keywords}

\section{Introduction}
Galaxy physical appearance (or morphology) and star formation rate (SFR) are two of the most common properties used to classify galaxies.
There is some linkage between the two such that frequently, we see that passive galaxies possess large galactic bulges, whereas star-forming galaxies are more discy in appearance \citep[e.g.][]{strateva2001,driver2006,bamford2009, bluck2014,morselli2017}. Quantifying whether these trends are causal or coincidental is required before we can fully understand what makes a galaxy stop forming stars.
 
There is also a very strong correlation between a galaxy's  SFR and its stellar mass, $M_{\star}$.
This correlation means that star-forming galaxies are confined to a narrow sequence \citep[with scatter of order $\sim0.3$ dex, see][and references within]{speagle2014} on the log(SFR) vs. log($M_{\star}$) plane, dubbed the star-forming main sequence \citep[SFMS;][]{noeske2007}.
This fundamental scaling relation covers several dex in stellar mass and describes a (mostly) linear increase in log(SFR) with $\log(M_{\star})$. This relation was in place early \citep{schreiber2015,leslie2020}, and while the sequence is tight, the physics of what drives the scatter in the SFMS (especially at high stellar masses) is of great interest.

Many recent works find that the SFMS relation is not linear across the entire range of stellar masses mapped by extragalactic surveys. Instead, it bends such that high-mass galaxies ($\log M_{\star}(\rm{M}_{\odot})\gtrsim10.5$ at $z=0$) possess lower SFRs than projected for their mass based on an extrapolation of the relation for lower-mass galaxies \citep[e.g.][]{noeske2007, bauer2013, whitaker2014, whitaker2015, schreiber2015, tomczak2016,leslie2020}. The reason for this decrease in SFR at high stellar masses is unknown, but at low redshifts is thought to be due to a combination of the effects of stellar mass, morphology, and environment \citep{erfanianfar2016}.
Indeed, various works have studied the link between main sequence bending and secular processes such as gas depletion due to environmental effects \citep[e.g.][]{gavazzi2015}, AGN feedback \citep[e.g.][]{mancuso_main_2016, brennan2017}, halo quenching \citep[e.g.][]{popesso2019}, or disc rejuvenation \citep[e.g.][]{mancini2019}.

The growth of a component that increases the stellar mass of a galaxy but not its SFR could also cause the observed decrease in galaxy specific SFR (sSFR) at high stellar masses.
For this reason, bulges have also been proposed as a morphological driver of SFMS bending \citep{wuyts2011, abramson2014, lang2014, whitaker2015, erfanianfar2016}. 
The growth of a dispersion-dominated bulge has also been linked to the cessation of star formation in a galaxy via a morphological quenching pathway \citep{martig2009}. In this manner, a disc may be stabilized against further fragmentation through the growth of a central mass concentration. However, this paradigm does not explain observations of  bulge-dominated galaxies residing in the highly star-forming region of the SFR vs. $M_{\star}$ diagram \citep[e.g.][]{wuyts2011, morselli2017, popesso2019}.

Bulges can form and grow via multiple pathways, including mergers \citep[e.g.][]{hopkins2010}, or in a secular manner \citep[e.g.][]{pfenniger1990}. Stellar bars are known to play an important role in bulge formation by driving gas into the central regions of galaxies \citep[e.g.][]{quillen1995}, resulting in starbursts \citep[e.g.][]{spinoso2017}, and contributing to central mass concentration growth \citep[e.g.][]{wang2012}. Given bars are disc phenomena, we expect the bulges formed by their influence to be rotation-supported by nature \citep[e.g.][]{bittner2020}.

But just how can a bulge grow in an actively star-forming galaxy without also quenching the galaxy? `Compaction' describes the growth of a bulge through the movement of galaxies around the main sequence plane through both internal and external processes \citep[e.g.][]{zolotov2015, tacchella2016}. Star-forming galaxies may propagate upwards to be above the SFMS line when an episode of gas infall is triggered (be that by mergers, counter-rotating streams, or violent disc instabilities). During this episode, gas is funnelled to the central regions of a galaxy, where it is used up in a burst of star formation \citep[e.g.][]{ellison2018}, fuelling the growth of central regions to a saturation point. After this starburst ceases, a galaxy will drop down onto the SFMS (or below) as the gas-depleted galaxy waits to become replenished again. The complex interplay between depletion and replenishment times determines the position of a galaxy on the SFR vs. $M_{\star}$ diagram today. In this manner, a galaxy will build up its bulge (and become more compact) through successive compaction events, whilst remaining on the SFMS. Importantly, the process of compaction sets no constraints on bulge kinematics.

One of the results of compaction should be a population of bulge-dominated galaxies that lie above the SFMS. \citep{morselli2017, popesso2019}. Some studies however, do not find this, \citep[e.g.][]{cook2020}, and rather attribute the bulge-dominated starbursting galaxies to poor bulge-disc decompositions, often complicated by mergers and interactions. These same works that suggest bulge growth as the cause of SFMS bending also report that this process is not sufficient to produce the amount of bending seen at high stellar masses \citep{popesso2019}. Indeed, main sequence bending has also been seen in populations of visually classified pure disc galaxies \citep{guo2015}. These studies suggest that a decrease in the SF activity of the disc is also required, and various environmental mechanisms including virial shock heating \citep[e.g.][]{birnboim2003, keres2005} or gravitational infall heating \citep[e.g.][]{dekel2008,khochfar2008} have been proposed to provide this additional star formation quenching. 

Whatever the cause of the SFMS bending, we do know that the scatter in the SFMS likely reflects a real diversity in star formation histories \citep{abramson2014, matthee2019}. 
Extending on this idea, we might also expect a variety of stellar kinematics, indicative of a variety of galaxy formation pathways. Previous work has shown a link between Hubble type and the spin parameter $\lambda_{Re}$ \citep{Cortese16,falcon-barroso2019, wang2020}, $V/\sigma$ \citep{vandesande2018}, and specific angular momentum \citep{Cortese16}, such that later-type spiral galaxies are more rotationally-supported than earlier-type spirals and S0s.
\citet{wang2020} extended on this idea by examining the link between galaxy visual morphology and position on the SFMS. A picture emerged in which galaxies lying on the SFMS were predominantly spirals with small bulges, while below the SFMS, galaxy kinematics depended on stellar mass \citep[though it should be noted that our own Milky Way violates this picture with a small bulge, but low SFR for its stellar mass e.g.][]{licquia2015}. \citet{wang2020} reported a strong mass dependence below the SFMS such that low-mass galaxies were `fast rotator' early-type galaxies, while high-mass ($M_{\star}>2\times10^{11}~\rm{M}_{\odot}$) galaxies were `slow rotator' spheroids.

A dichotomy at $z=0$ between star-forming, disc-dominated galaxies and passive, bulge-dominated galaxies is apparent. What is unclear however, is the order of these processes. Can a bulge form in a star-forming galaxy (and does it have a role in the quenching of star formation), or is bulge build-up the realm of passive galaxies?

In this paper, we investigate kinematic trends across the SFMS with IFS data, comparing galaxy spin parameters both on and off the SFMS. 
For this sort of analysis, we will benefit from the number statistics that the two largest IFS surveys to date can provide, and so we combine data from both the Sydney-AAO Multi-object Integral field spectrograph \citep[SAMI;][]{croom2012} galaxy survey and the Mapping Nearby Objects at APO \citep[MaNGA;][]{bundy2015} galaxy survey. Given that the target selection of these two surveys differ, we are able to probe more of the galaxy parameter space, and compare whether or not trends seen in one data set persist between the two.
To enable the best comparison possible, we measure kinematic properties between the two surveys using a homogeneous set of structural parameters, SFR measurements and stellar mass indicators.

This paper is organised as follows: in Section~\ref{data} we describe the SAMI and MaNGA IFS surveys, along with the homogeneous structural parameters used to calculate kinematic measurements. We also describe the IFS sample, kinematic measurement and corrections, along with the definition of the main sequence line used. In Section~\ref{results} we present the results, and in Section~\ref{discussion} we discuss the implications of our findings. Throughout this paper we employ a $\Lambda$CDM cosmology, with $\Omega_{m}=0.3$, $\Omega_{\lambda}=0.7$, $H_{0}= 70~ \rm{km}~\rm{s}^{-1}~\rm{Mpc}^{-1}$ and a \citet{chabrier2003} IMF.

\section{Data and methods}
\label{data}
\subsection{The SAMI galaxy survey}
The SAMI galaxy survey is an IFS survey on the Anglo-Australian Telescope (AAO) that observed 3068 galaxies from 2013--2018 \citep{croom2012}. 
SAMI uses 13 fused fibre hexabundles \citep{bland-hawthorn2011,bryant2014} with a high (75\%) fill factor. Each bundle contains 61 fibres of $1.6^{\prime\prime}$ diameter resulting in each integral field unit (IFU) having a diameter of $15^{\prime\prime}$. The IFUs, as well as 26 sky fibres, are plugged into pre-drilled plates using magnetic connectors. SAMI fibres are fed to the double-beam AAOmega spectrograph \citep{sharp2015}, which allows a range of different resolutions and wavelength ranges. 
The SAMI Galaxy survey employs the 570V grating at 3750--5750 \AA~giving a resolution of R=1810 ($\sigma = 70.4~\rm{km}~\rm{s}^{-1}$) at 4800 \AA, and the 1000R grating from 6300--7400 \AA~giving a resolution of R=4260 ($\sigma = 29.6~\rm{km}~\rm{s}^{-1}$) at 6850 \AA~\citep{scott2018}. 83\% of galaxies in the SAMI target catalogue have coverage out to 1$R_{e}$ \citep{bryant2015}.

The SAMI survey is comprised of a sample drawn from the GAMA equatorial regions \citep{bryant2015}, and an additional sample of eight clusters \citep{owers2017}. SAMI Data Release 3 \citep[DR3;][]{croom2021} contains observations of 3068 galaxies and is the final data release of the SAMI survey. 
SAMI DR3 includes observations spanning $0.04<z<0.128$ and $7.42<\log M_{\star}[\rm{M}_{\odot}] < 11.89$ (corresponding to an $r$-band magnitude range of $18.4 < m_{r} < 12.1$), with environments ranging from underdense field regions to extremely overdense clusters. 

SAMI DR3 galaxy cubes are provided for use, along with an array of maps data products. All data products have spaxel size of $0.5^{\prime\prime}~\rm{spaxel}^{-1}$, and the average seeing FWHM is $\sim2^{\prime\prime}$. Here, we employ the two-moment Gaussian  line of sight velocity distribution (LOSVD) stellar kinematic maps \citep{vandesande2017}, including rotational velocity, and velocity dispersion ($\sigma$) maps. We use the adaptively binned maps, in which spaxels are binned to a signal-to-noise (S/N) of 10 using the Voronoi binning code of \citet{cappellari2003}. The S/N is calculated from the flux and variance spectra of each spaxel as the median across the entire blue wavelength range \citep{scott2018}, and spaxels with S/N $>$10 are not binned. 

\subsection{The MaNGA galaxy survey}

The MaNGA Galaxy Survey is an IFS
survey that observed $>$10,000 galaxies from 2014--2020 \citep{bundy2015, drory2015}. It is an SDSS-IV project \citep{blanton2017}, employing the 2.5m telescope at Apache Point Observatory \citep{gunn2006} and BOSS spectrographs \citep{smee2013}, which have continuous wavelength coverage from 3600--10300 \AA~ at R$\sim2700$ ($\sigma\sim70$ $\rm{km}~\rm{s}^{-1}$). MaNGA’s target galaxies were chosen to include a wide range of galaxy masses and colours, over the redshift range $0.01 < z < 0.15$. The Primary+ sample \citep{yan2016, wake2017} contains
galaxies with spatial coverage out to $\sim 1.5 R_{e}$ for $\sim66\%$ of
the total sample, and the remainder (dubbed the Secondary
sample) are observed out to $\sim 2.5 R_{e}$, generally at higher redshifts than the Primary+ sample.
SDSS-IV data release 15 \citep[DR15;][]{aguado2019} contains 4621 unique galaxies, selected in the range $7.9<\log M_{\star}[\rm{M}_{\odot}] < 12.1$, (corresponding to $18.1 < m_{r} < 11.6$), and a range of field environments,
observed and reduced by the MaNGA Data Reduction Pipeline \citep{law2015}. Derived properties are produced by the MaNGA Data Analysis Pipeline \citep[DAP;][]{westfall2019}, provided as a single data cube per galaxy \citep{yan2016a}. MaNGA's spaxel size matches that of SAMI, at $0.5^{\prime\prime}~\rm{spaxel}^{-1}$, and the average seeing conditions throughout the survey were such that the $r$-band PSF FWHM is $\sim2.5^{\prime\prime}$.

We employ the two-moment LOSVD stellar velocity and dispersion maps using the Voronoi binning scheme to ensure each bin reaches a target S/N of 10. We also apply the velocity dispersion correction provided to account for MaNGA instrumental dispersion \citep[see][]{westfall2019}.

\subsection{Star formation rates and stellar masses}
In this analysis we wish to compare trends in the spin parameter $\lambda_{Re}$ with current star formation activity in galaxies. Given there may be observational biases that are unaccounted for between the two surveys, we report trends in SAMI and MaNGA data separately. However, to determine a robust star-forming main sequence line, we wish to be able to place the two surveys on a homogeneous SFR-$M_{\star}$ plane.

For this reason, we match both SAMI DR3 and MaNGA DR15 to the \textit{GALEX}-Sloan-\textit{WISE} legacy catalogue 2 \citep[GSWLC-2;][]{salim2016, salim2018} using a sky match with maximum separation of $2^{\prime\prime}$. GSWLC-2 provides UV--optical--mid-infrared (IR) SED-derived stellar masses and SFRs for 659,229
galaxies within the SDSS footprint and $z < 0.3$, with photometry provided by \textit{GALEX}, SDSS, and the Wide-Field Survey Explorer (\textit{WISE}). We utilise the GSWLC-X2 catalogue, which uses the deepest \textit{GALEX} photometry available (selected from the shallow ‘all-sky’, medium-deep, and deep catalogues) for a source in the SED fit. SED fitting was performed using the Code Investigating GALaxy Emission \citep[CIGALE;][]{noll2009, boquien2019}, which constrains SED fits with IR luminosity, which they
term SED+LIR fitting. 

3901 MaNGA galaxies have matches to the GSWLC-2, and 1832 SAMI galaxies. Unfortunately many of the galaxies lost belong to the SAMI cluster sample, though we note that four clusters have GSWLC-2 coverage.

\subsection{Structural parameters}
To enable a comparison between SAMI and MaNGA kinematic quantities, we require the structural parameters used to define the apertures to be identical. Indeed, a small change in aperture size can result in an appreciable difference in  $\lambda_{Re}$ values for a given galaxy. 
For this reason, we match both surveys to the NASA-Sloan Atlas \citep[NSA;][]{blanton2011}, and use the elliptical Petrosian values for effective radius ($R_{e}$), axis ratio ($b/a$, which we use to define the ellipticity, $\epsilon$, as $\epsilon = 1 - b/a$), photometric galaxy position angle ($\phi$), and the S\'{e}rsic index (n) from a single S\'{e}rsic fit. As MaNGA's targeting catalogue was the NSA, all galaxies have these values available. 1831 SAMI galaxies with GSWLC-2 data also have counterparts in the NSA. 

\subsection{Defining a star-forming main sequence}
\begin{figure}
\centering
\begin{subfigure}{0.49\textwidth}
\includegraphics[width=\textwidth]{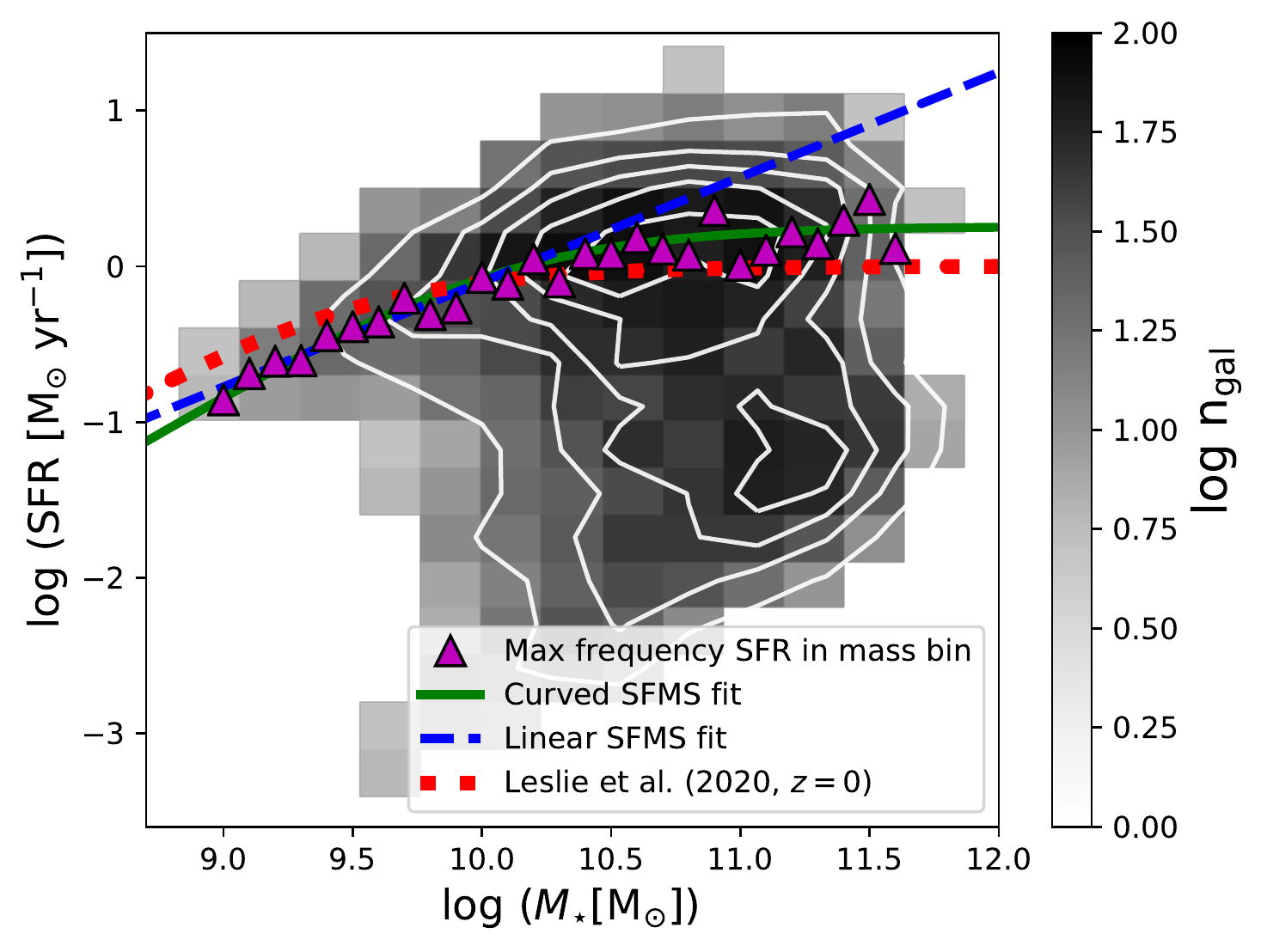}  
\end{subfigure}
\caption{A 2D histogram of the number of galaxies ($\rm{n}_{\rm{gal}}$) in bins of stellar mass and star formation rate for the combined SAMI and MaNGA sample. Contours of the the overall distribution of the sample are overlaid in white. 
The SFMS line is fit to GSWLC-2 galaxies that match the overall redshift distribution of SAMI and MaNGA, and magenta markers denote the peak of the SFR distribution for that mass bin. The SFMS line of Equation~\ref{sfms} is shown in green and the linear SFMS line fit of Equation~\ref{eqn2} is shown in blue. For comparison, the SFMS fit of \citet{leslie2020} extrapolated to $z=0$ is shown in red. }
\label{MS_def}
\end{figure}

We define the SFMS line for the SAMI and MaNGA galaxies used in this work by fitting a curve to the points at which the number density is highest in the SFR vs. $M_{\star}$ diagram in bins of stellar mass over the mass range $9<\log M_{\star}[\rm{M}_{\odot}]<11.7$. For each bin of stellar mass, we simply determine the peak of a histogram of SFRs. 
In order to increase number statistics, whilst still fitting to the overall SAMI and MaNGA galaxy distribution, we fit this curve to the GSWLC-2 sample over the redshift range of the SAMI and MaNGA samples. We also weight the GSWLC-2 galaxy sample such that the overall redshift distribution of the SAMI and MaNGA samples is also matched. This increases the number of galaxies used in the fit from $\sim5700$ in the SAMI and MaNGA sample to $\sim403,000$. 
In this manner, the main sequence line naturally bends at high stellar masses, as shown in Figure~\ref{MS_def}. 
We fit the functional form of the main sequence curve definition introduced in \citet{leslie2020} \citep[which is based on that of][]{lee2015}:
\begin{align}
    \log(\langle SFR[\rm{M}_{\odot}~\rm{yr}^{-1}] \rangle) &= S_{0} - a_{1}t - \log_{10}\left(1+\frac{10^{M'_{t}}}{10^{M}}\right),\\
    M'_{t} &= M_{0} - a_{2}t,
\end{align}
where $M$ is $\langle \log(M_{\star}/\rm{M}_{\odot})\rangle$, $M'_{t}$ is the turnover mass, and $t$ is the age of the Universe in Gyr. For star-forming galaxies, \citet{leslie2020} find $S_{0}=2.97^{+0.08}_{-0.09}$, $M_{0}=11.06^{+0.15}_{-0.16}$, $a_{1}=0.22^{+0.01}_{-0.01}$, and $a_{2}=0.12^{+0.03}_{-0.02}$. We use these values, along with $t=13.5$ Gyr, and fit a curve to the star-forming galaxies (defined arbitrarily as those with $\log(SFR) > 0.704 \times \log(M_{\star}) - 8.21$) in the GSWLC-2 using \textsc{scipy's optimize.curvefit} package.  
The best fit main sequence line for the SAMI and MaNGA galaxies is:
\begin{equation}
    \log(SFR[\rm{M}_{\odot}~\rm{yr}^{-1}]) = 0.256 - \log_{10} \left(1 + \frac{10^{10.064}}{10^{M}}\right), 
    \label{sfms}
\end{equation}
where $M$ is as defined above.

We note that Equation~\ref{sfms} deviates at high stellar masses towards slightly higher SFRs compared to the \citet{leslie2020} curve, as shown in Figure~\ref{MS_def}; similar SFMS behaviour is also reported in \citet{thorne2020}. The \citet{leslie2020} SFMS relation is derived from 3GHz radio continuum imaging of the COSMOS field, and is extrapolated below $z\sim0.3$. The high-mass objects observed at low-redhift in this work are not present in the \citet{leslie2020} sample, and we speculate this is the reason behind the discrepancies at the high-mass end.

As a comparison, we also fit a linear main sequence line to investigate any biases introduced by the assumption that the main sequence bends at high stellar masses. We use only low-mass galaxies in the linear main sequence fit, (where there is no obvious deviation from a straight line, see Figure~\ref{MS_def}) within the mass range $9.0 < \log M_{\star}[\rm{M}_{\odot}] < 10.0$. The best fit straight line to the GSWLC-2 galaxies scaled to match the redshift distribution of the combined SAMI and MaNGA sample is:

\begin{equation}
\log(SFR[\rm{M}_{\odot}~\rm{yr}^{-1}]) = (0.674\times M) - 6.836,
\label{eqn2}
\end{equation}
 where $M$ is as defined above. The results of this paper using a linear SFMS line are presented in Appendix~\ref{appendix}.

We define the quantity $\Delta~\rm{MS}$ as the difference in SFR from the prediction of the SFMS curve of Equation~\ref{sfms} for a galaxy of the same mass. 
 \begin{figure}
\centering
\begin{subfigure}{0.49\textwidth}
\includegraphics[width=\textwidth]{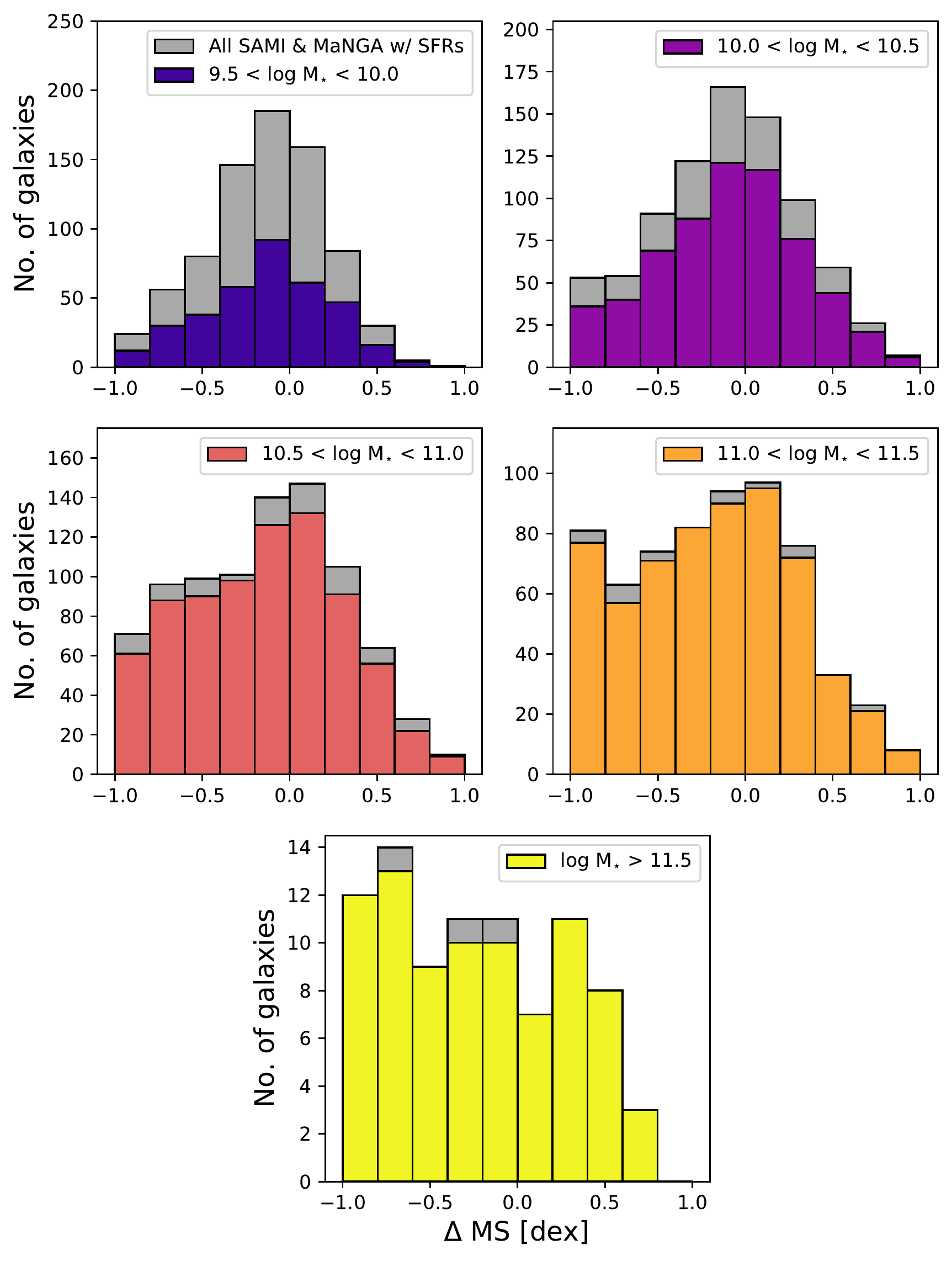}  
\end{subfigure}
\caption{The effects of the kinematic sample selection cuts on the parent sample of SAMI and MaNGA galaxies with GSWLC-2 SFRs. Each panel is a histogram of the distribution of the distance of a galaxy to the main sequence line ($\Delta~\rm{MS}$) in five mass bins, with the parent sample in grey, and the final sample of galaxies used for kinematic analysis after quality cuts were made in colour. The majority of galaxies rejected from the kinematic analysis are low-mass, and this is mostly due to low continuum S/N. }
\label{samples}
\end{figure}

\subsection{$\lambda_{Re}$ measurement}
\label{kins}
Following \citet{emsellem2007} and \citet{emsellem2011}, we define the spin parameter approximation, $\lambda_{Re}$, as the flux-weighted ratio of ordered to disordered motion within a galaxy:
\begin{equation}
\lambda_{R} = \frac{\langle R  |V|\rangle}{\langle R  \sqrt{V^{2} + \sigma^{2}}\rangle} = \frac{\sum_{i=0}^{N_{spx}} F_{i}R_{i}|V_{i}|}{\sum_{i=0}^{N_{spx}} F_{i}R_{i}\sqrt{V^{2}_{i} + \sigma^{2}_{i}}},
\label{lamReqn}
\end{equation}
where $F$ is the flux, $V$ the stellar rotational velocity, and $\sigma$ the stellar velocity dispersion of the $ith$ spaxel. In the same manner as \citet{Cortese16} and \citet{vandesande2017}, we define $R$ as the semi-major axis of an ellipse on which spaxel $i$ lies. We chose to use the intrinsic radius rather than the circular projected radius as it follows the galaxy light profile more accurately. We note that while this is the same technique used for SAMI galaxies by \citet{vandesande2017}, the values of $\epsilon$, $R_{e}$, and $\phi$ used to define the ellipse within which $\lambda_{Re}$ is calculated are different. The reason for this difference is that we wish to compare SAMI and MaNGA measurements in as close a manner as possible, and hence used the same catalogue (the NSA) for structural measurements of galaxies for both surveys.
This discrepancy results in a small scatter of order $\sim0.05$ in $\lambda_{Re}$ measurements (though importantly, no offset) between the $\lambda_{Re}$ values from \citet{vandesande2017} and those reported in this work.

At this point, some cuts were also applied to the SAMI and MaNGA data to ensure only galaxies with reliable kinematics were included in the $\lambda_{Re}$ catalogue.
In both samples, we removed galaxies with $R_{e}$ less than the HWHM PSF of the observation. 
For MaNGA, we also removed galaxies for which more than 20\% of spaxels within an ellipse of semi-major axis 1$R_{e}$ were masked. The masking could be the result of flags introduced in the data reduction process (as the DAP velocity and $\sigma$ masks were applied to the maps prior to analysis), or we also masked all spaxels where the corrected $\sigma<50~\rm{km}~\rm{s}^{-1}$, as \citet{westfall2019} suggests that this is the lower limit for which dispersion measurements can be trusted when S/N$>10$. 

SAMI kinematic quality cuts are described in Section 3.2.6 of \citet{vandesande2017}, and involve a relative $\sigma$ cut such that bad spaxels are defined as those with $\sigma_{\rm{error}} > \sigma~\times~0.1 + 25~\rm{km}~\rm{s}^{-1}$. We keep the same quality cuts as \citet{vandesande2017}, and reject any galaxy with $>25\%$ bad spaxels from the following analysis. In addition, we removed any galaxies for which $R_{e}$ is greater than the aperture size ($\sim16\%$ of the sample), to avoid the need for aperture corrections, and those that were flagged as having unreliable kinematics in the SAMI DR3 kinematics catalogue.

In Figure~\ref{samples}, we show histograms of the combined SAMI and MaNGA parent sample with GSWLC-2 SFRs (grey histograms), and the final kinematic sample used in this analysis after all cuts are made (coloured histograms). Each panel of Figure~\ref{samples} represents a mass bin used in this work. Unsurprisingly, the greatest number of galaxies are lost from the low stellar mass bins, mostly due to poor continuum S/N within the galaxy. 
Our final samples are representative and highly complete (84\%) for $\log M_{\star}[\rm{M}_{\odot}] > 10$, though the completeness drops significantly (to 48\%) for $\log M_{\star}[\rm{M}_{\odot}] < 10$. While we still cover the entire range of SFRs of interest for our analysis, we recommend caution in extrapolating our findings to the entire low mass population. In summary,
897 SAMI galaxies and 2392 MaNGA galaxies have reliable $\lambda_{Re}$ measurements. 

\subsection{Inclination and seeing corrections}
\begin{figure}
\centering
\begin{subfigure}{0.49\textwidth}
\includegraphics[width=\textwidth]{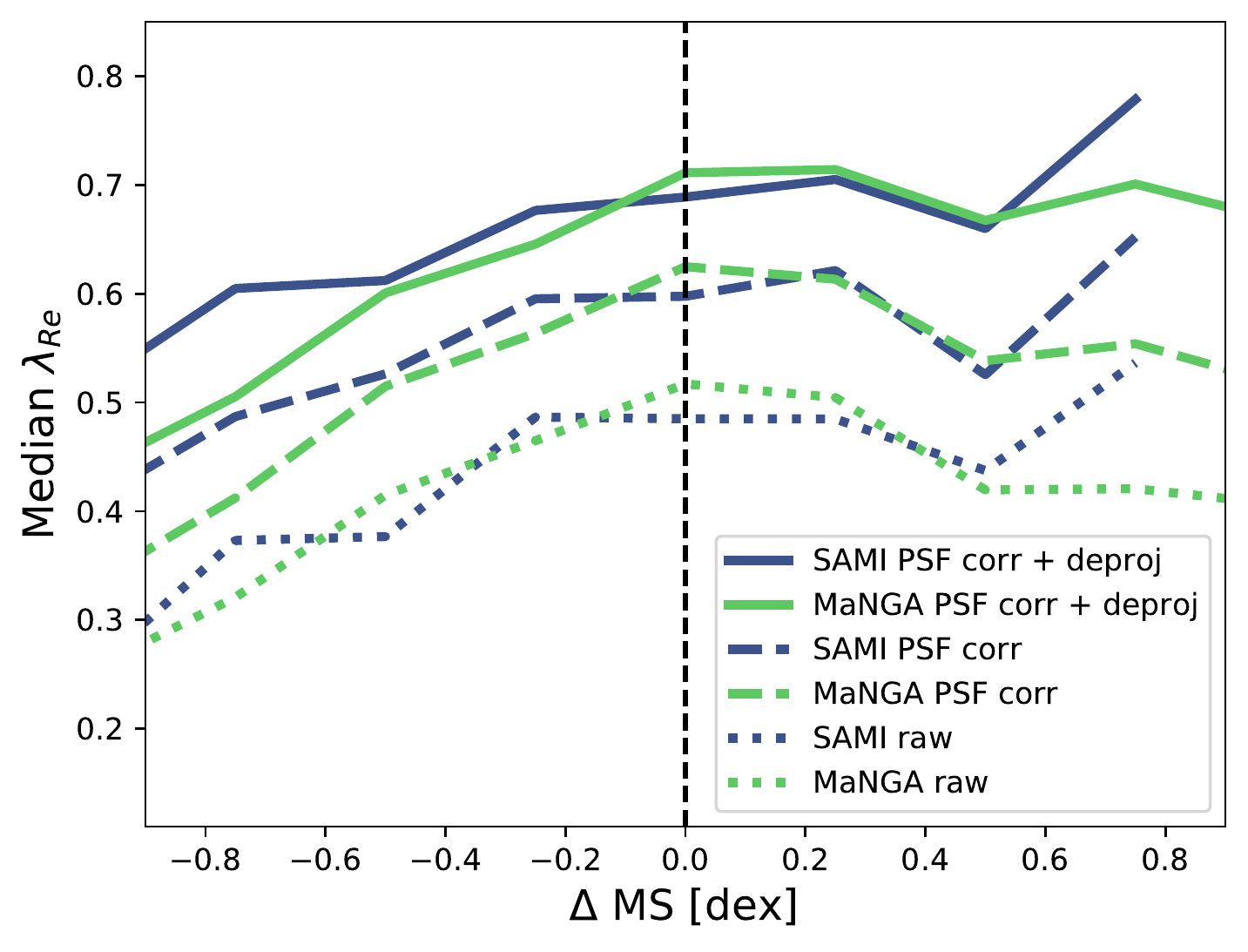}  
\end{subfigure}
\caption{The effect of the PSF and inclination corrections on the spin parameter $\lambda_{Re}$. Navy lines indicate SAMI median values, and green MaNGA. Dotted lines denote the raw median $\lambda_{Re}$ values, dashed lines are after the PSF-correction of \citet{harborne2020}, and solid lines are median $\lambda_{Re}$ values in bins of $\Delta~\rm{MS}$ after PSF-correction and deprojection.}
\label{lam_PSF_deproj}
\end{figure}

Measurement of the $\lambda_{Re}$ parameter is influenced by both the FWHM of the PSF of the observation and the galaxy inclination angle \citep[e.g.][]{cappellari2016, graham2018}, hence we attempt to account for both of these effects.  
Given the difference in average seeing conditions between the SAMI (FWHM $\sim2^{\prime\prime}$) and MaNGA (FWHM $\sim2.5 ^{\prime\prime}$) surveys, it is essential to apply a seeing correction so that we may facilitate as close a comparison in kinematic properties as possible. There are several recent examples of seeing corrections for IFS data in the literature \citep[e.g.][]{graham2018, chung2020, harborne2020}.
We decide to apply the seeing correction of \citet{harborne2020}\footnote{http://github.com/kateharborne/kinematic\_corrections}, due in part to its ease of application to different IFS survey datasets. Briefly, the corrections of \citet{harborne2020} take the S\'{e}rsic index of the galaxy and FWHM of the IFS observation and provide a value for $\lambda_{Re}$ that is corrected for seeing. Given the MaNGA PSF is on average $\sim0.5^{\prime\prime}$ greater than that of SAMI, the PSF corrections affect the MaNGA data more. 
Figure~\ref{lam_PSF_deproj} shows the increase in median $\lambda_{Re}$ after both PSF correction \citep[an increase of $\sim0.1$, in line with][]{graham2018} and deprojection corrections are applied as a function of distance from the main sequence ($\Delta~\rm{MS}$) for SAMI (navy blue lines) and MaNGA (green lines) galaxies. Although the corrections change the absolute value of the median $\lambda_{Re}$, the overall shape of the curves are preserved, meaning that the relative ordering of the spin parameter values will not change greatly when kinematic corrections are applied.

 \begin{figure}
\centering
\begin{subfigure}{0.49\textwidth}
\includegraphics[width=\textwidth]{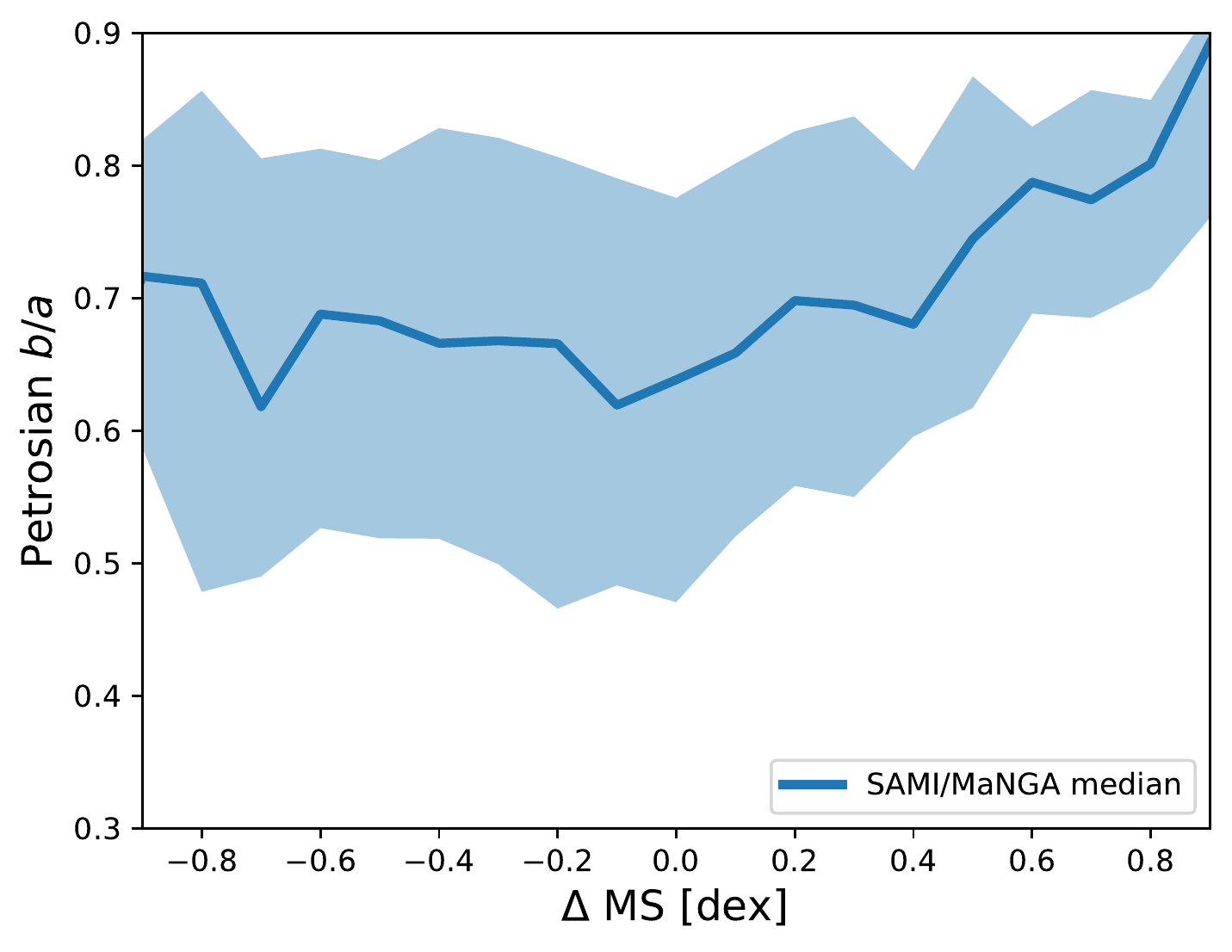}  
\end{subfigure}
\caption{Petrosian axial ratio, $b/a$, of the combined SAMI/MaNGA kinematic sample for bins of $\Delta$MS. The shaded region denotes the 25$^{th}$ and 75$^{th}$ percentiles of the distribution. Galaxies above the main sequence are on average more round than those on or below.}
\label{inc_MS}
\end{figure}

Figure~\ref{inc_MS} shows that for the combined SAMI and MaNGA sample, there is a dependence on galaxy axis ratio ($b/a$) with $\Delta~\rm{MS}$ such that galaxies above the main sequence are rounder than those on the main sequence (assuming that $b/a$ indicates inclination and not intrinsic shape). 
As pointed out by \citet{wang2020}, without an inclination correction the raw rotational stellar velocity and $\sigma$ values propagate to artificially low $\lambda_{Re}$ values above the main sequence, making these galaxies appear more dispersion-dominated than they actually are. 

Many deprojection corrections exist in the literature ranging from a simple $1/\sqrt{\epsilon}$ \citep[e.g.][]{Cortese16}, to more complicated functions \citep[e.g.][]{falcon-barroso2019}. We chose the correction of \citet{emsellem2011}, as implemented by \citet{delmoral-castro2020}:
\begin{align}
    \lambda^{deproj}_{R} &= \frac{\lambda_{R}}{\sqrt{C^{2} - \lambda_{R}^{2}(C^{2}-1)}}; \\
    C &= \frac{\sin i}{\sqrt{1 - \beta \cos^{2}i}};\\
    \cos i &= \sqrt{\frac{\left(b/a\right)^{2} - q_{0}^{2}}{1 - q_{0}^{2}}}
\label{deprojEqn}
\end{align}
where $b/a$ is the axis ratio of the galaxy, and $q_{0}$ is the intrinsic axial ratio of an edge-on galaxy. As we are interested primarily in galaxies on or near the star-forming main sequence, we choose $q_{0}=0.2$, as used in \citet{Cortese16} for galaxies with a clear disc component. The anisotropy parameter, $\beta$, varies slightly with Hubble type, but we use $\beta=0.3$, which is appropriate for disc galaxies \citep[derived from Table B.1 of][]{kalinova2017}.

We note that previous studies have found variation in both the $q_{0}$ and $\beta$ parameters with galaxy morphology \citep[e.g.][]{cappellari2007, chemin2018}. 
We tested the difference between SAMI $\lambda_{Re,~deproj}$ values using fixed $q_{0}$ and $\beta$ and those where $q_{0}$ and $\beta$ varied with galaxy morphology obtained from the catalogue of \citet{Cortese16}. We found very little difference between the two methods, with the maximum $\Delta \lambda_{Re,~deproj}$ of $\sim0.03$. Importantly, there are no trends in median $\lambda_{Re,~deproj}$ with $\Delta~\rm{MS}$.
In addition, van de Sande et al. (\textit{MNRAS, in prep.}), show that there is very little offset ($\Delta \lambda_{Re}<0.05-0.1$) in deprojected $\lambda_{Re}$ measures between using the simplistic assumptions presented above and a more detailed approach as described by \citet{cappellari2007}. 

Given the similarity between the fixed and morphology-based deprojection values coupled with the fact that we do not have a homogeneous morphology catalogue for both the SAMI and MaNGA samples, we stick with the assumption of $q_{0}=0.2$ and $\beta=0.3$. 

Figure~\ref{lam_PSF_deproj} shows that while the overall shape of the $\lambda_{Re}$ distribution as a function of $\Delta~\rm{MS}$ remains similar, on average, MaNGA $\lambda_{Re}$ values are slightly higher ($\Delta \lambda_{Re,~PSF~corr + deproj}\sim$0.05) than SAMI at $\Delta~\rm{MS}=0$. There could be multiple reasons for this discrepancy, one of which being simply a difference in sample selection. That said, the locus of the main sequence should be well sampled by both surveys. A difference in the median S/N may also be the result of MaNGA sampling more of the disc regions of galaxies that are missed by SAMI. While poorly sampled galaxies are removed from the kinematic samples and we always measure $\lambda$ out to 1$R_{e}$, if MaNGA is sampling slightly more spaxels per galaxy on average than SAMI, this may result in a slightly higher median $\lambda_{Re}$ measurement. Finally, another possible reason for the $\lambda_{Re}$ discrepancy may be the way in which stellar velocity and $\sigma$ were derived between surveys. While SAMI broadens their spectra to that of the templates used for a continuum fit, MaNGA fits at the native resolution, then applies a dispersion correction after fitting to account for instrumental dispersion effects. Both of these methods produce velocity and $\sigma$ measurements that convey the astrophysical Doppler broadening, though it is possible that the differing techniques result in slight differences between the resultant derived velocity and $\sigma$ measurements.

There is currently no galaxy that is observed in both SAMI DR3 and MaNGA DR15 releases, but we note that if this changes in the future (indeed, \citet{law2020} found 74 galaxies in common between the internal MaNGA Product Launch 10 (MPL-10) and SAMI DR2), a detailed analysis into any discrepancies between velocity and $\sigma$ measures will be extremely informative. Additionally, performing the analysis of this work on simulated SAMI and MaNGA kinematic data will give insight into the origin of any small differences seen in the kinematics between the two surveys.
  

\section{Results}
\label{results}
\subsubsection*{$\lambda_{Re}$ on the SFMS}
After performing the various sample cuts described in Section~\ref{data}, the SAMI sample spans $0.01<z<0.11$, $9.5<\log M_{\star}[\rm{M}_{\odot}]<11.8$, $17.8<m_{r}<12.1$, and the MaNGA sample $0.01<z<0.15$, $9.5<\log M_{\star}[\rm{M}_{\odot}]<12.1$, $17.5<m_{r}<11.6$.

\begin{figure}
\centering
\begin{subfigure}{0.49\textwidth}
\includegraphics[width=\textwidth]{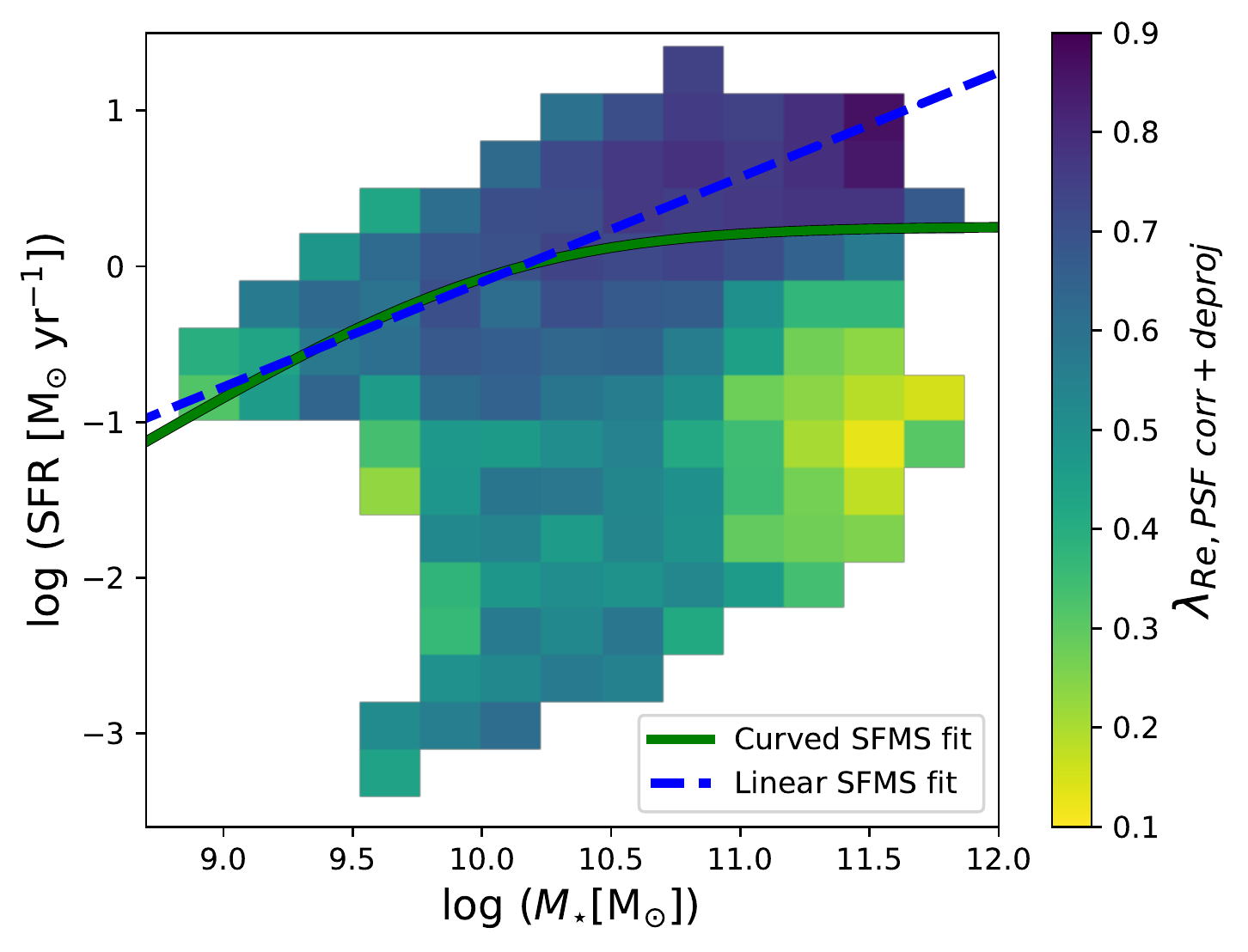}  
\end{subfigure}
\caption{A 2D histogram of median $\lambda_{Re}$ for bins of stellar mass and star formation rate for the combined SAMI and MaNGA sample.
The SFMS line of Equation~\ref{sfms} is shown in green, and for comparison, the linear main sequence line fit to low-mass galaxies of Equation~\ref{eqn2} is shown in blue.}
\label{lamr_MS}
\end{figure}

In Figure~\ref{lamr_MS}, we present a 2D histogram of the SFR vs. $M_{\star}$ plane for both the SAMI and MaNGA samples with bins coloured by the average PSF-corrected and deprojected $\lambda_{Re}$ values. The main sequence line defined in Equation~\ref{sfms} is shown in green. In line with Croom et al. (\textit{MNRAS, submitted}), overall trends are readily visible: the passive, high-mass galaxies are chiefly dispersion-dominated systems, and the rotation-dominated systems populate the main sequence line regions of the plot. Interestingly, the range of $\lambda_{Re}$ values is greatest at the highest stellar masses: high-mass galaxies are both the most rotation-dominated \textit{and} the most dispersion-dominated galaxies in the local Universe.

 \begin{figure*}
\centering
\begin{subfigure}{0.8\textwidth}
\includegraphics[width=\textwidth, trim= 0 7cm 0 0, clip]{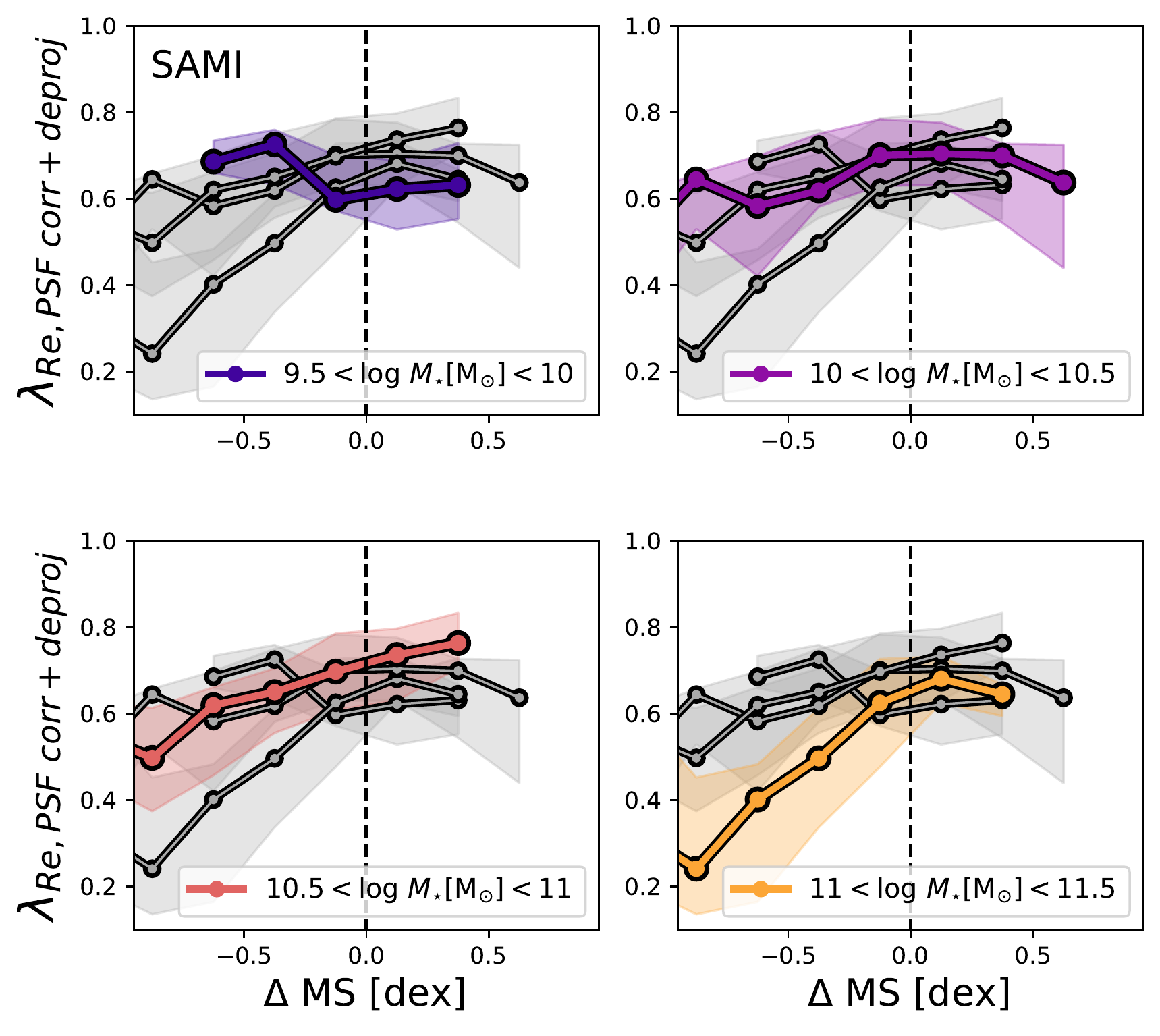}  %
\end{subfigure}
\caption{Median $\lambda_{Re}$ for SAMI galaxies as a function of distance from the SFMS line of Equation~\ref{sfms} ($\Delta~\rm{MS}$) in bins of stellar mass. Each panel highlights a different mass bin, with all other mass bins shown in grey for comparison. Shaded regions denote the 25$^{th}$ and 75$^{th}$ percentiles for each mass bin. The vertical dashed line denotes the locus of the main sequence. In general, the SFMS is populated by discy galaxies, with little change in $\lambda_{Re}$ within $\pm$ 1 dex of the SFMS for all but the most massive galaxies.}
\label{results2_SAMI}
\end{figure*}

 \begin{figure*}
\centering
\begin{subfigure}{0.8\textwidth}
    \includegraphics[width=\textwidth]{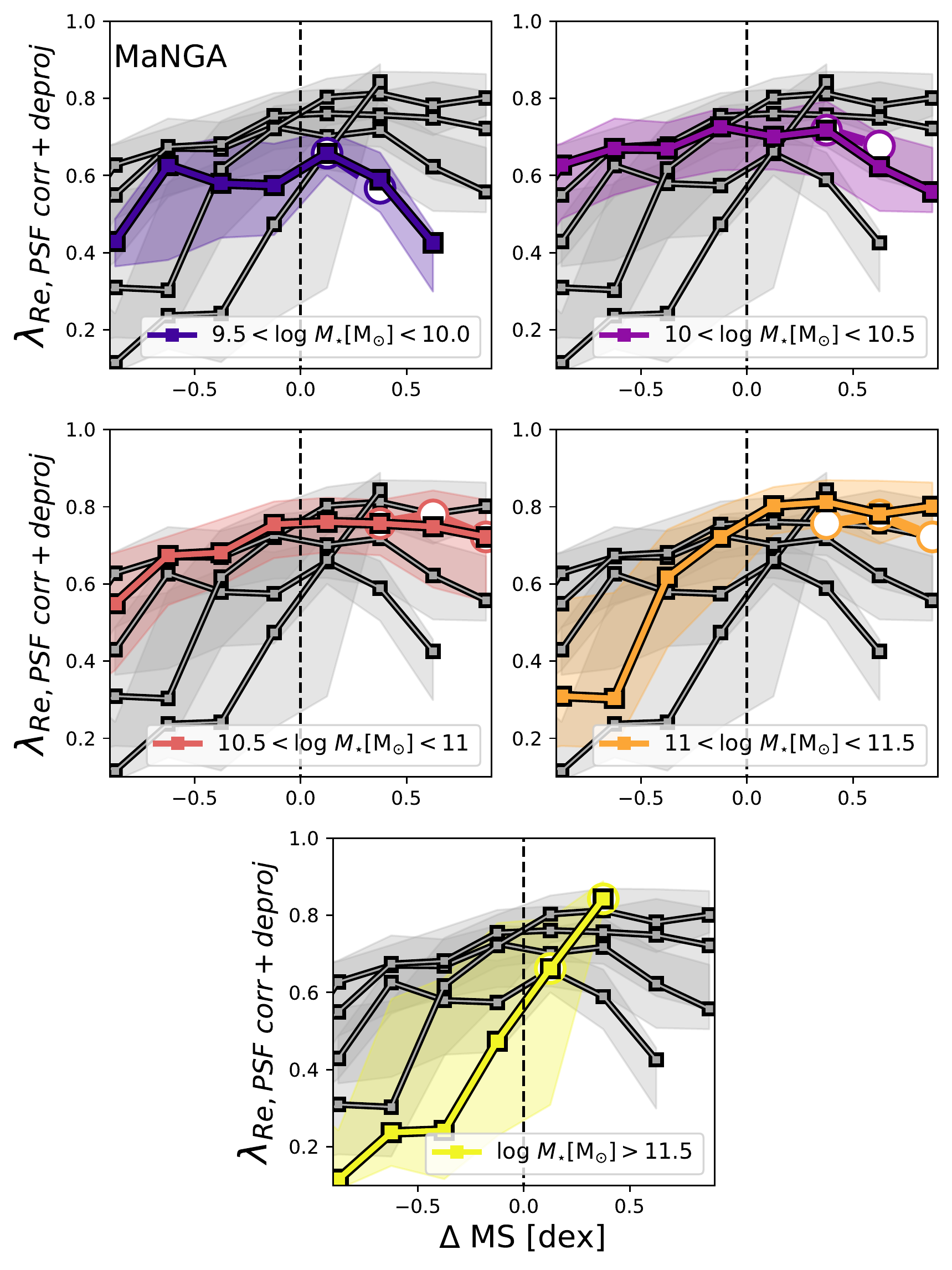}  
\end{subfigure}
\caption{Median $\lambda_{Re}$ for MaNGA galaxies as a function of distance from the SFMS line of Equation~\ref{sfms} ($\Delta~\rm{MS}$) in bins of stellar mass. Each panel highlights a different mass bin, with all other mass bins shown in grey for comparison. Shaded regions denote the 25$^{th}$ and 75$^{th}$ percentiles for each mass bin. Solid lines depict the full MaNGA kinematic sample, and dotted lines with white circles are the MaNGA sample with obvious interactions removed from the highest two $\Delta~\rm{MS}$ bins. The extra high-mass bin for MaNGA galaxies is due to the greater number of high-mass galaxies in this sample.}
\label{results2_MaNGA}
\end{figure*}

 We plot the median $\lambda_{Re}$ as a function of $\Delta~\rm{MS}$ for the SAMI kinematic sample in Figure \ref{results2_SAMI}, and for the MaNGA kinematic sample in Figure~\ref{results2_MaNGA}, where only bins that contain five or more galaxies are displayed. As a comparison, we plot the same parameters in Figures~\ref{A1} and ~\ref{A2}, but calculating $\Delta~\rm{MS}$ using the linear fit to low-mass galaxies from Equation~\ref{eqn2}. 
 In both Figures, each panel highlights a mass bin, with all others shown for comparison in grey. Shaded regions denote the 25$^{th}$ and 75$^{th}$ percentiles. Due to increased sample statistics at the high-mass end, there is an extra high-mass bin for the MaNGA sample that is not present for the SAMI data. 
For both the SAMI and MaNGA results in Figures~\ref{results2_SAMI} and~\ref{results2_MaNGA}, we see that at the locus of the main sequence ($\Delta~\rm{MS}=0$), the median $\lambda_{Re}$ value is high. Galaxies on the main sequence are rotationally-supported systems, in line with previous photometric \citep[e.g.][]{wuyts2011, morselli2017} and spectroscopic \citep{oh2020,wang2020} structural studies.
Both SAMI and MaNGA suggest an increase of $\lambda_{Re}$ with stellar mass for main sequence galaxies. Specifically, the median $\lambda_{Re}$ increases from $\sim$0.65 for $9.5 < \log M_{\star}[\rm{M}_{\odot}]< 10$ to $\sim$0.75 for  
$10.5 < \log M_{\star}[\rm{M}_{\odot}]< 11$ at $\Delta~\rm{MS}=0$ for MaNGA galaxies.
Overall, we see a hint of mass dependence for $\lambda_{Re}$ such that the slope of the median $\lambda_{Re}$ as a function of $\Delta~\rm{MS}$ becomes steeper at higher stellar masses. 

Above the main sequence, the SAMI sample does not show any significant change in the median value of the stellar spin parameter. However, this sample does not probe beyond $\Delta~\rm{MS}\sim$0.4 dex. Conversely, MaNGA allows us to reach $\Delta~\rm{MS}\sim$0.8 dex where, at least for stellar masses $9.5 < \log M_{\star}[\rm{M}_{\odot}]< 11$, we find marginal evidence for a decrease in $\lambda_{Re}$ in very strongly star-forming galaxies. While intriguing, this decrease is only marginally significant, and given the tendency of tidal interactions triggering starbursts, potentially more indicative of disturbances in the stellar velocity field than gradual thickening of the disk or build-up of a dispersion-dominated stellar component. 

Indeed, if we remove the 27 galaxies that clearly show signs of gravitational interaction in their SDSS optical images from the highest two bins of $\Delta~\rm{MS}$ for the MaNGA sample, the decrease in stellar spin at high $\Delta~\rm{MS}$ reduces somewhat. 
In Figure~\ref{results2_MaNGA}, solid lines depict the full MaNGA kinematic sample, and dotted lines are the MaNGA sample with obvious interactions removed from the highest two $\Delta~\rm{MS}$ bins in the right panel. All disturbed SAMI galaxies were already removed from the sample when the quality control cuts were applied.
Figure~\ref{results2_MaNGA} confirms that especially for stellar masses $10 < \log M_{\star}[\rm{M}_{\odot}]< 11$, the median $\lambda_{Re}$ value flattens out slightly above the main sequence. 
We note that we removed only the most obviously interacting systems whose SDSS images showed extreme warping from tidal interaction. There are likely many interacting systems of varying degrees of tidal disruption still remaining within the MaNGA kinematic sample. 

Below the main sequence, the picture emerging is slightly different. For galaxies with stellar masses $9.5 < \log M_{\star}[\rm{M}_{\odot}]< 10.5$, stellar spin seems to remain roughly constant up to $\sim$1 dex below the locus of the main sequence. At higher stellar masses, both SAMI and MaNGA seem to suggest a steepening of the $\Delta~\rm{MS}$--$\lambda_{Re}$ relation so that with increasing mass, low stellar spin galaxies become more frequent closer to the locus of the main sequence.

\section{Discussion}
\label{discussion}
 \begin{figure*}
\centering
\begin{subfigure}{0.99\textwidth}
\includegraphics[width=\textwidth]{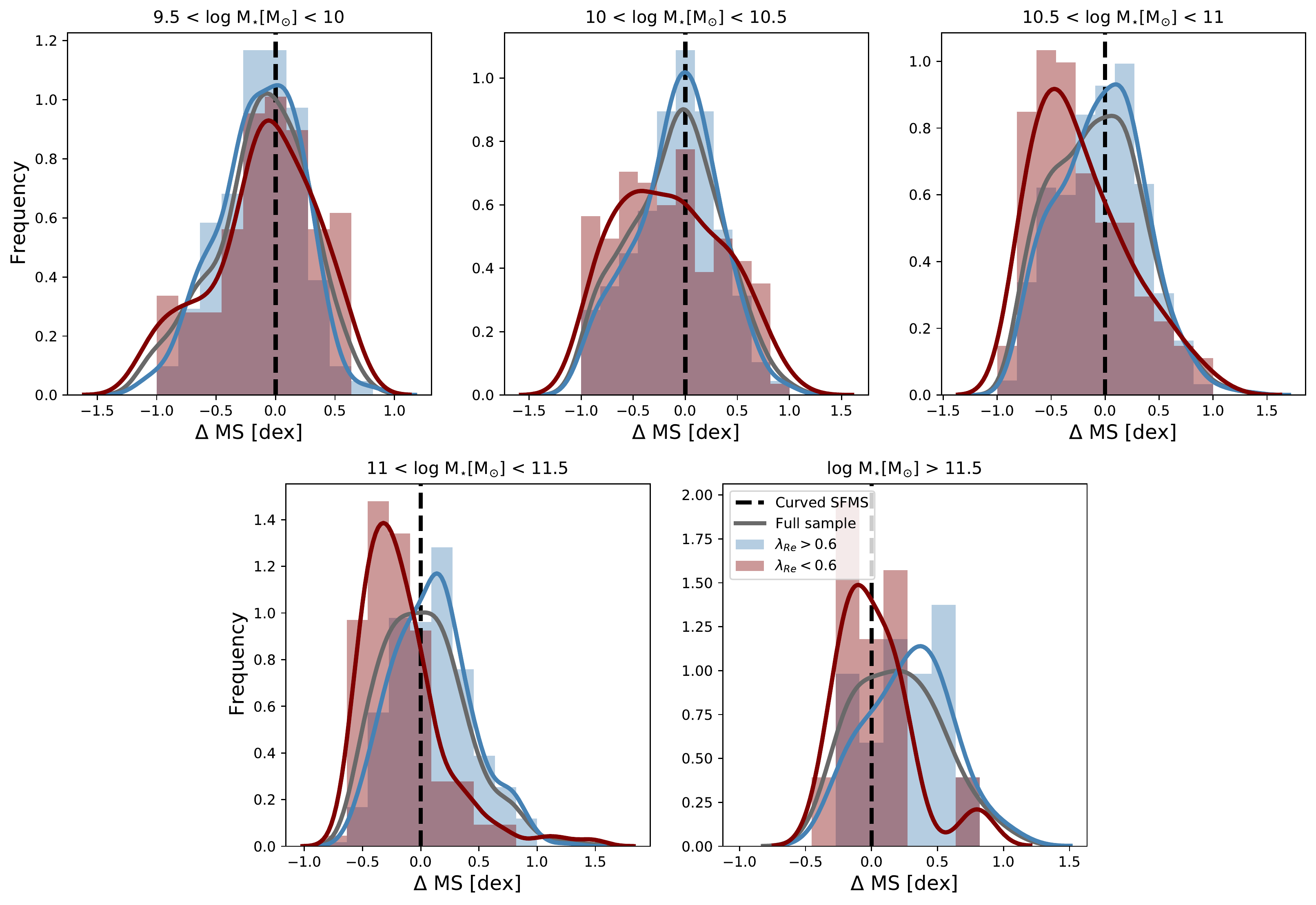}  
\end{subfigure}
\caption{Histograms and their associated kernel density estimate (KDE) plots of the combined SAMI and MaNGA kinematic sample (grey), galaxies with $\lambda_{Re}>0.6$ (blue), and $\lambda_{Re}<0.6$ (red) in bins of stellar mass for $-1<\Delta~\rm{MS}<1$. $\Delta~\rm{MS}=0$ is shown in black. 
At low stellar masses, there is no deviation away from the main sequence.
At higher stellar masses, low $\lambda_{Re}$ galaxies deviate towards lower sSFRs, while high $\lambda_{Re}$ galaxies move above the main sequence line. We interpret these trends as evidence that more dispersion-dominated galaxies populate the `bending' region of the SFMS.}
\label{bendingcause}
\end{figure*}

\subsection{A discy main sequence}
We firstly discuss trends seen for star-forming galaxies located on the main sequence.
The most striking observation from Figures~\ref{results2_SAMI} and~\ref{results2_MaNGA} is that apart from the highest stellar mass bin of $\log M_{\star}[\rm{M}_{\odot}]>11.5$, all galaxies on the main sequence ($\Delta~\rm{MS}=\pm0.25$ dex) possess $\lambda_{Re}$ values indicative of discy galaxies ($0.6 < \lambda_{Re,~PSF~corr+deproj} < 0.8$). 
As there is a strong link between the $\lambda_{Re}$ spin parameter and the intrinsic shape of a galaxy \citep[e.g.][]{foster_sami_2017}, we can therefore infer that galaxies on the main sequence are intrinsically flattened and axisymmetric discs. Apart from perhaps the highest mass bins, galaxies on the main sequence are as rotationally-supported and intrinsically flattened as they will get; the main sequence is populated by the disciest galaxies.

For the MaNGA galaxies in Figure~\ref{results2_MaNGA}, we see a small mass dependence at $\Delta~\rm{MS}=0$ such that apart from the highest mass bin (within which dispersion-dominated structures may be beginning to dominate) there is the trend that on average, higher-mass galaxies have greater values of $\lambda_{Re}$ than low mass. 
\citet{catinella2006} show that the rotation curves of high-mass galaxies reach their peaks at shorter disc scale-lengths than low-mass dwarfs, making it more likely that the flat region of their rotation curves are contained within 1$R_{e}$. Given the $\lambda_{Re}$ metric effectively normalises galaxy spin by stellar mass (thanks to the $\sigma$ in the denominator of Equation~\ref{lamReqn}), on average, the peak velocity of a galaxy's rotation curve should be contained within the 1$R_{e}$ aperture of high-mass galaxies more frequently than for their low-mass counterparts. Hence, the resultant $\lambda_{Re}$ value will be greater.
This observation may explain the mass dependence seen at the locus of the main sequence.
 
\subsection{Classical bulge growth above the main sequence?}
Recent photometric studies of galaxies above the main sequence report that the bulge-to-total ratio ($B/T$) increases such that starbursting galaxies are more bulge-dominated than their neighbours on the main sequence \citep{morselli2017, popesso2019}. These authors find that starbursting galaxies possess highly star-forming central regions, which from concentration measures they infer are resultant from the growth of classical bulges. We are able to test this theory from a kinematic standpoint.

While we do observe a slight reduction in median $\lambda_{Re}$ above the main sequence, this reduces when we remove the 27 galaxies from the MaNGA sample that are obviously interacting according to their SDSS colour images. These interacting galaxies will likely be highly dispersion-supported due to the random motions of stars induced by merger activity. Of course interactions act to decrease the rotational support of a galaxy whilst often inducing starburst activity, however these motions are not necessarily indicative of classical bulge growth. Galaxy interactions may therefore be artificially lowering the median spin parameter value above the main sequence.

We re-examine trends in the $\lambda_{Re}-\Delta~\rm{MS}$ relation of Figure~\ref{results2_MaNGA} above the main sequence once interacting galaxies are removed. For all but the highest and lowest stellar mass bins, the median $\lambda_{Re}$ curves flatten somewhat such that they are similar both on and above the main sequence, and these values are for that of dynamically cold, discy systems. Our results show that there is very little decrease in $\lambda_{Re}$ for the majority of non-interacting systems above the main sequence, and from this we imply that we do not see evidence of classical bulge growth in this regime. 

The finding that dispersion-dominated bulges are not growing above the main sequence for the majority of the galaxy population does not preclude a compaction scenario. Rather, it is constrained such that the episode of gas infall must occur in a manner so as not to disrupt the rotationally-supported nature of the inner regions of a galaxy. We speculate that the gas infall event that triggers a central burst of star formation must be ordered. An investigation into the ordered and random motions of gas in the central regions of galaxies above the main sequence should reveal just how turbulent the gas infall episode is.

We note that the SAMI and MaNGA samples do not probe the extreme starbursting galaxy population. Our results are statistically significant up to $+0.8$ dex above the SFMS, while photometric studies such as \citet{morselli2017} report $B/T$ trends up to $+1$ dex above. Figure 5 of \citet{morselli2017} shows that the steepest increase in $B/T$ above the main sequence occurs between $+0.5$ and $+1$ dex above the SFMS. Given these trends, we should still expect to see some evidence of dispersion-dominated structure growth in the highest two $\Delta~\rm{MS}$ bins of our kinematic results. We do see some evidence for a decrease in spin parameter, but we attribute this to a small number of interacting galaxies, rather than large-scale classical bulge growth. 

One such explanation for the discrepancy between the kinematic results presented here and photometric results from the literature is from the photometric decomposition technique. \citet{cook2020} showed that the structural decomposition technique used can affect $B/T$ measurements across a large mass range. From a careful structural decomposition of a relatively small sample of galaxies, \citet{cook2020} found a monotonic decrease in $B/T$ above the main sequence for all but the lowest-mass galaxies in their sample. They attributed the observed differences to spurious structural measurements stemming from the limited model validation available for large (SDSS-sized) catalogues of bulge-disc decompositions. Indeed, mergers and interacting galaxies are traditionally very difficult to fit with simple bulge+disc models \citep[e.g.][]{mezcua2014}.
Our work agrees qualitatively with that of \citet{cook2020}: we do not find evidence of a population of starburst galaxies with systematically higher $B/T$ in the local Universe. 

A caveat to this work is the spatial resolution of the IFS observations; it is possible that these galaxies on and above the main sequence do not have bulges large enough to be seen in the stellar kinematics. The PSF of SAMI and MaNGA are $\sim2^{\prime\prime}$ and $\sim2.5^{\prime\prime}$ respectively, which both correspond to 2.0 kpc at the median redshifts of the kinematic samples used in this work of $z=0.05$ and $z=0.04$. 
Dispersion-dominated bulges significantly smaller than 2 kpc may be washed out through beam-smearing effects. While this should not be a problem for higher-mass galaxies, classical bulges located in lower-mass galaxies can indeed possess sub-kpc bulge effective radii \citep{gadotti2009}.

These results can be linked to the structural growth and morphological transformation within galaxies in the context of star formation. 
Given that we see no growth of dispersion-supported structure on the SFMS, and yet passive galaxies host such structures (especially at high stellar masses), we may say something about the link between galaxy quenching and morphological transformation via dispersion-dominated bulge growth. Our results are consistent with two scenarios: the first where initial quenching must take place \textit{before} morphological transformation, and the second where if these two processes are concurrent, then the timescales differ such that morphological transformation occurs more slowly than quenching \citep[or at least the galaxy moving off the main sequence; e.g.][]{cortese2019}.
We are not in a position to say which scenario is occurring, but Croom et al. (\textit{MNRAS, submitted}) takes a different approach in attempting to explain the formation of S0 galaxies via a combination of photometric concentration measures and kinematic dispersion parameterisation. In this manner, they find that S0 formation can be explained via a simple disk fading model taking into account progenitor bias. These results may provide clues about bulge growth in the wider galaxy population.

 \subsection{Classical bulge growth below the main sequence?}
Figures~\ref{results2_SAMI} and~\ref{results2_MaNGA} show a steepening of the $\lambda_{Re}$--$\Delta~\rm{MS}$ relation with increasing stellar mass below the SFMS. The reason behind this steepening is unclear: while it seems to be revealing an increase in dispersion-supported structure dominance, it could also be the result of an upwards scatter in SFRs due to the inherent unreliability of SFR indicators at low sSFRs.

Separating star-forming and passive galaxies becomes increasingly difficult at higher stellar masses. As can be seen from Figure~\ref{MS_def}, the clear bi-modality of star-forming and passive populations seen between $10.5 < \log M_{\star}[\rm{M}_{\odot}]<11.0 $ diminishes at higher stellar masses. Coupled with a bending of the SFMS towards lower SFRs, it becomes difficult to determine where the main sequence is sampling star-forming galaxies, and where the green valley begins. Indeed, works that define a SFMS through Gaussian mixture modelling have increasing difficulty fitting two Gaussians (one for the star-forming population and one for the passive population) at high stellar masses \citep[e.g.][]{popesso2019}. Whether this blending of populations in the SFR plane is physical or the result of unreliable SFR indicators at low sSFR is unknown. If the latter, then we might expect some passive galaxies to artificially inhabit the lower portion of main sequence regions. This effect would be strongest at high stellar masses for a curved main sequence, as it is in these regions that the main sequence line deviates to lower sSFRs. The observed steepening of the $\lambda_{Re}$--$\Delta~\rm{MS}$ relation may be explained by a portion of passive galaxies (with dominant dispersion-supported structure) contaminating the $\lambda_{Re}$ measures below the SFMS. Indeed, the steepening of the $\lambda_{Re}$--$\Delta~\rm{MS}$ relation with mass practically disappears if we use a linear fit to the main sequence.

If the observed steepening of the $\lambda_{Re}$--$\Delta~\rm{MS}$ slope is real, then this would suggest that the mechanisms acting on high- and low-mass galaxies as they become more passive are different: one produces passive galaxies with similar disc structure as when they were on the main sequence, while the other must dramatically alter the kinematics of a galaxy. The obvious mechanism that will destroy or thicken a disk is mergers. Interestingly, the vast majority of slow rotator galaxies possess high stellar masses \citep[e.g.][]{emsellem2007, vandesande2017a, graham2018, vandesande2020, wang2020}. It is tempting to speculate that the reason for the $\lambda_{Re}$ steepening in high-mass galaxies only  may be that either the mergers required to create them only occur in high-mass galaxies, or perhaps the processes of mass build up as the result of mergers differ with stellar mass \citep[e.g.][]{robotham2014}. Both of these processes must begin while the galaxy is still on the SFMS. 

One subject that this work does not touch on is the effect of environment on the degree of dispersion support within galaxies as a function of their sSFR.
Hence, an exciting avenue for follow-up work on this topic is through exploring trends with centrals vs. satellite galaxies.

\subsection{The cause of main sequence bending}
\label{bending_MS}
Many works propose the growth of bulges as the driver of main-sequence bending \citep[e.g.][]{abramson2014, popesso2019}. Already we see a hint in Figure~\ref{results2_MaNGA} that the highest-mass galaxies (the regime in which we expect the greatest deviation from a linear main sequence) are more dispersion-dominated.
We are in a unique position to test this theory from a kinematic standpoint by examining whether we see any differences in the $\Delta~\rm{MS}$ values of high- and low-$\lambda_{Re}$ galaxies. 

We split the combined SAMI and MaNGA sample between $-0.8<\Delta~\rm{MS}<0.8$ into low ($\lambda_{Re}<0.6$) and high ($\lambda_{Re} >0.6$) $\lambda_{Re}$ sub-samples. We note here that the low $\lambda_{Re}$ sample does not consist solely of dispersion-dominated systems, rather they are simply \textit{more} dispersion-supported than the high $\lambda_{Re}$ systems. There are also trends present with stellar mass such that higher-mass galaxies are more likely to possess greater dispersion support. This means that there will be a greater number of high-mass galaxies in the low-$\lambda_{Re}$ sample, and lower-mass galaxies in the high-$\lambda_{Re}$ sample. In Figure~\ref{bendingcause}, we plot the distribution of $\Delta~\rm{MS}$ for low $\lambda_{Re}$ (red line) and high $\lambda_{Re}$ (blue line) galaxies as a function of distance from the curved SFMS line defined in Equation~\ref{sfms}. As a comparison, we plot the distribution of the overall sample in grey. The locus of the SFMS is shown by a black dashed line.

At low stellar masses we see that the $\Delta~\rm{MS}$ distribution is very similar for all values of $\lambda_{Re}$, though the low-$\lambda_{Re}$ systems begin to deviate above $\log M_{\star}[\rm{M}_{\odot}] =10$, and at high masses are preferentially located below the SFMS line. 
Similarly, above $\log M_{\star}[\rm{M}_{\odot}]=11$, the high-$\lambda_{Re}$ systems begin to deviate above the overall $\Delta~\rm{MS}$ distribution.
At high-mass, systems with greater dispersion dominance preferentially populate regions below the SFMS line (however it is defined), whilst rotation-dominated systems sit above. We interpret these trends as evidence that the `bending' region of the SFMS is populated by galaxies of greater dispersion support -- high-mass galaxies with greater dispersion support are more likely to possess lower SFRs than their more rotationally-dominated counterparts. 

Our findings suggest that dispersion-dominated bulges are already present in massive galaxies on the main sequence. This is not surprising, given that the existence of visually classified early-type (i.e. possessing a prominent bulge component) star-forming spirals has been known since the establishment of the Hubble morphological sequence. That said, the growth of a dispersion dominated bulge is not the only possible cause of a decrease in $\lambda_{Re}$: disc thickening will also decrease $\lambda_{Re}$. When our results are coupled with photometric work highlighting the redistribution of stars towards central regions below the main sequence however \citep[e.g.][]{morselli2017, popesso2019}, they are sufficient to expect that at least some of the $\lambda_{Re}$ decrease is due to bulge growth. 

It is very tempting to push the interpretation of our results further and wonder if they provide direct evidence of a physical link between lower SFRs and the growth of dispersion-dominated structure in high-mass galaxies.
The morphological quenching argument of \citet{martig2009} suffices in explaining the lower SFRs seen in high-mass galaxies with greater dispersion support. These galaxies possess lower SFRs because their bulges are large enough that they have begun to stabilise galaxy discs against further star formation.
A similar explanation was put forward by both \citet{whitaker2015} and  \citet{erfanianfar2016} to explain the morphology dependence on the scatter in the main sequence, and a flatter main sequence for galaxies with high S\'{e}rsic index respectively. 
It is also possible that the lower sSFR is due to the growth of a non-star-forming component that adds to the stellar mass of a galaxy without increasing its SFR. In this case, the growth of a bulge and the cessation of star formation do not need to be linked. Whatever the cause, we are left with an intriguing hint of the role of morphology in regulating a galaxy's star formation. We can certainly conclude that the bending of the SFMS at high stellar masses is coincident with a population of galaxies that possess classical bulges.

\section{Summary \& Conclusions}

We search for evidence of kinematic transformation in galaxies on the SFMS by examining the link between galaxy SFR and stellar kinematics from IFS observations. Combining the might of the SAMI and MaNGA IFS galaxy surveys, we calculate the spin parameter, $\lambda_{Re}$, in a homogenised manner for 3289 galaxies. Our main results are:

\begin{enumerate}
    \item \textbf{Galaxies on the SFMS possess $\lambda_{Re}$ values indicative of intrinsically flattened discs.} There is a small mass trend such that higher-mass galaxies appear to have higher $\lambda_{Re}$ values than lower-mass galaxies, which we expect is due to the peak of low-mass galaxy velocity fields being more likely to occur outside the 1$R_{e}$ aperture used in this work. 
    
    For the highest stellar mass bin ($\log M_{\star}[\rm{M}_{\odot}]>11.5$), we see a population of galaxies on the SFMS that possess a small dispersion-dominated bulge component (and possibly some contribution from a thickened disc).\\
    
    \item \textbf{No decrease in $\lambda_{Re}$ above the SFMS.} Once interacting galaxies are removed, $\lambda_{Re}$ measurements up to $+0.8$ dex above the SFMS are consistent with those on the SFMS for the majority of galaxies (though we see marginal evidence that this may not hold true for the lowest-mass galaxies of $9.5 < \log M_{\star}[\rm{M}_{\odot}]< 10$), from which we conclude that there is no growth of dispersion-dominated galaxy components while a galaxy is in a starburst phase. If compaction is occurring in highly star-forming galaxies, it cannot be contributing to classical bulge growth.\\
    
    \item \textbf{A decrease in $\lambda_{Re}$ below the SFMS for high-mass ($\log M_{\star}[\rm{M}_{\odot}] > 11$) galaxies.} One possibility for the decrease in median $\lambda_{Re}$ below the SFMS may be that the SFR indicator is unreliable at low sSFRs, scattering some green valley galaxies to higher SFRs than they should be. If the trend is real however, then quenching mechanisms must differ between high- and low-mass galaxies: low-mass galaxies are quenching without structure growth, while some mechanism is acting to both quench a galaxy \textit{and} dramatically adjust the stellar kinematics at $\log M_{\star}[\rm{M}_{\odot}] > 11$. The likely culprit is gravitational interactions. \\
    
    \item \textbf{Evidence for a tantalising phenomenological connection between the bending of the SFMS and an increase in galaxy dispersion support.} Lower $\lambda_{Re}$ galaxies are preferentially located on or below the SFMS line for $\log M_{\star} [\rm{M}_{\odot}] > 10.5$. More rotationally-supported systems ($\lambda_{Re}>0.6$) better follow a linear SFMS line. The bending of the SFMS is primarily due to the fact that lower $\lambda_{Re}$ galaxies start dominating the galaxy budget of the SFMS at high stellar masses, which we speculate is evidence for the growth of classical bulges.\\
    
\end{enumerate}

Our results indicate that bulge growth is occurring in high-mass galaxies on and just below the SFMS to some degree.
In addition, we see evidence that the growth of a dispersion-dominated bulge is linked to the bending of the SFMS at high stellar masses. While extremely promising, we note that further investigation is still required to precisely identify the link between the SFMS bending and an increase in dynamical pressure support.
Despite our observations, bulge growth is minor for the majority of galaxies on the SFMS. Given that most extremely massive passive galaxies are slow rotators, we find that extra bulge growth is still required once a galaxy has quenched to produce the red and dead S0s observed in the local Universe today.

\section{Acknowledgements}
The SAMI Galaxy Survey is based on observations made at the Anglo-Australian Telescope. The Sydney-AAO Multi-object Integral field spectrograph (SAMI) was developed jointly by the University of Sydney and the Australian Astronomical Observatory. The SAMI input catalogue is based on data taken from the Sloan Digital Sky Survey, the GAMA Survey and the VST ATLAS Survey. The SAMI Galaxy Survey is supported by the Australian Research Council Centre of Excellence for All Sky Astrophysics in 3 Dimensions (ASTRO 3D), through project number CE170100013, the Australian Research Council Centre of Excellence for All-sky Astrophysics (CAASTRO), through project number CE110001020, and other participating institutions. The SAMI Galaxy Survey website is http://sami-survey.org/.
LC is the recipient of an Australian Research Council Future Fellowship (FT180100066) funded by the Australian Government.
JvdS acknowledges support of an Australian Research Council Discovery Early Career Research Award (project number DE200100461) funded by the Australian Government.
NS acknowledges support of an Australian Research Council Discovery Early Career Research Award (project number DE190100375) funded by the Australian Government and a University of Sydney Postdoctoral Research Fellowship.
Parts of this research were conducted by the Australian Research Council Centre of Excellence for All Sky Astrophysics in 3 Dimensions (ASTRO 3D), through project number CE170100013. 
JJB acknowledges support of an Australian Research Council Future Fellowship (FT180100231).
JBH is supported by an ARC Laureate Fellowship and an ARC Federation Fellowship that funded the SAMI prototype.
AMM acknowledges support from the National Science Foundation under Grant No. 2009416.
M.S.O. acknowledges the funding support from the Australian Research Council through a Future Fellowship (FT140100255).

\section*{Data Availability}
The SAMI data presented in this paper are available from Astronomical Optics’ Data Central service at: https://datacentral.org.au/. The MaNGA data are available at: https://www.sdss.org/dr15/manga/manga-data/data-access/.

\appendix 
\section{Linear main sequence}
\label{appendix}
\begin{figure*}
\centering
\begin{subfigure}{0.8\textwidth}
\includegraphics[width=\textwidth, trim= 0 7cm 0 0, clip]{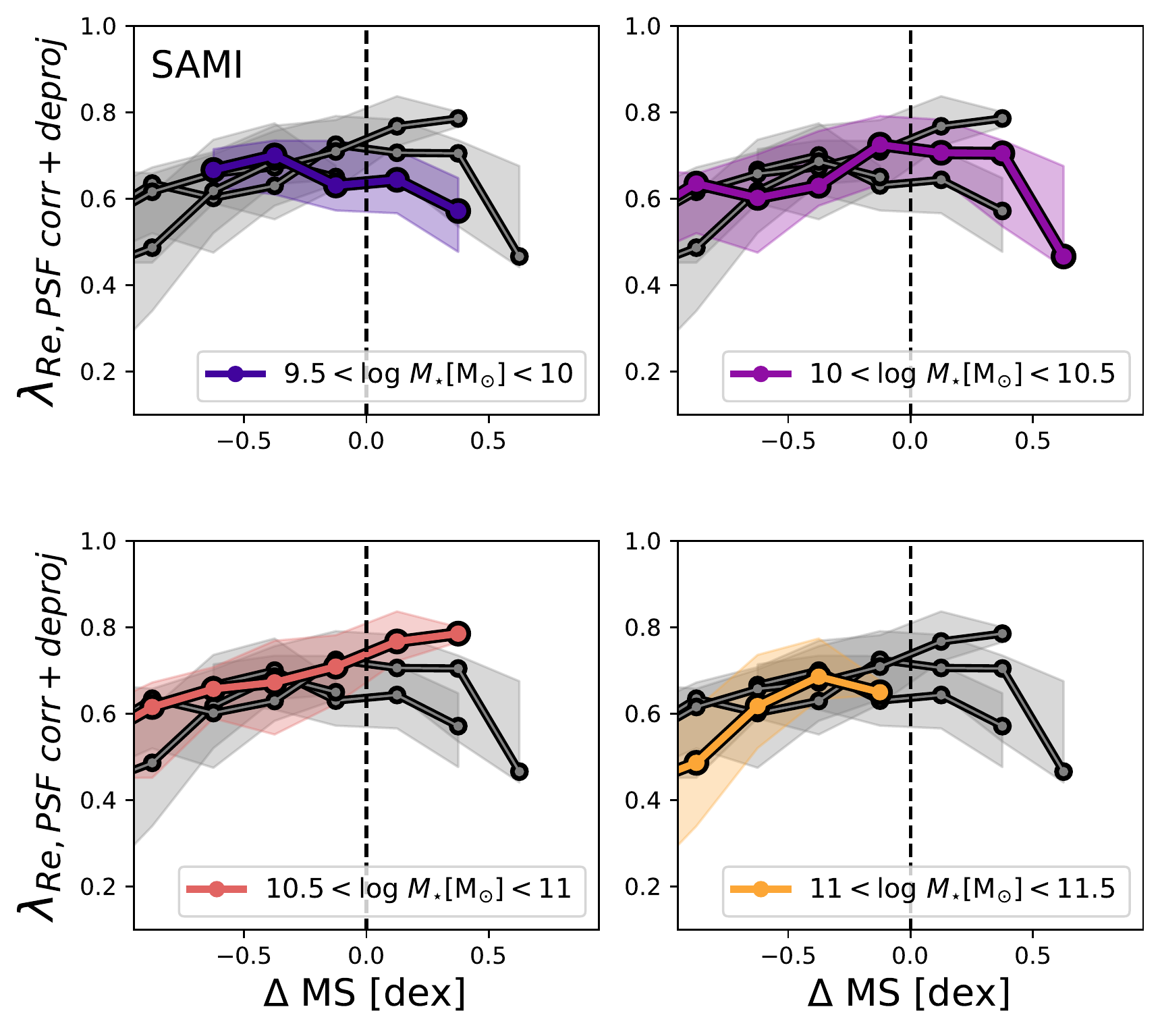}  
\end{subfigure}
\caption{Same as Figure~\ref{results2_SAMI}, but using the linear SFMS line Equation~\ref{eqn2} to calculate $\Delta~\rm{MS}$. Given there are fewer galaxies above the linear SFMS line, we are not able to probe as far above the main sequence as with the curved line.}
\label{A1}
\end{figure*}

\begin{figure*}
\centering
\begin{subfigure}{0.8\textwidth}
    \includegraphics[width=\textwidth]{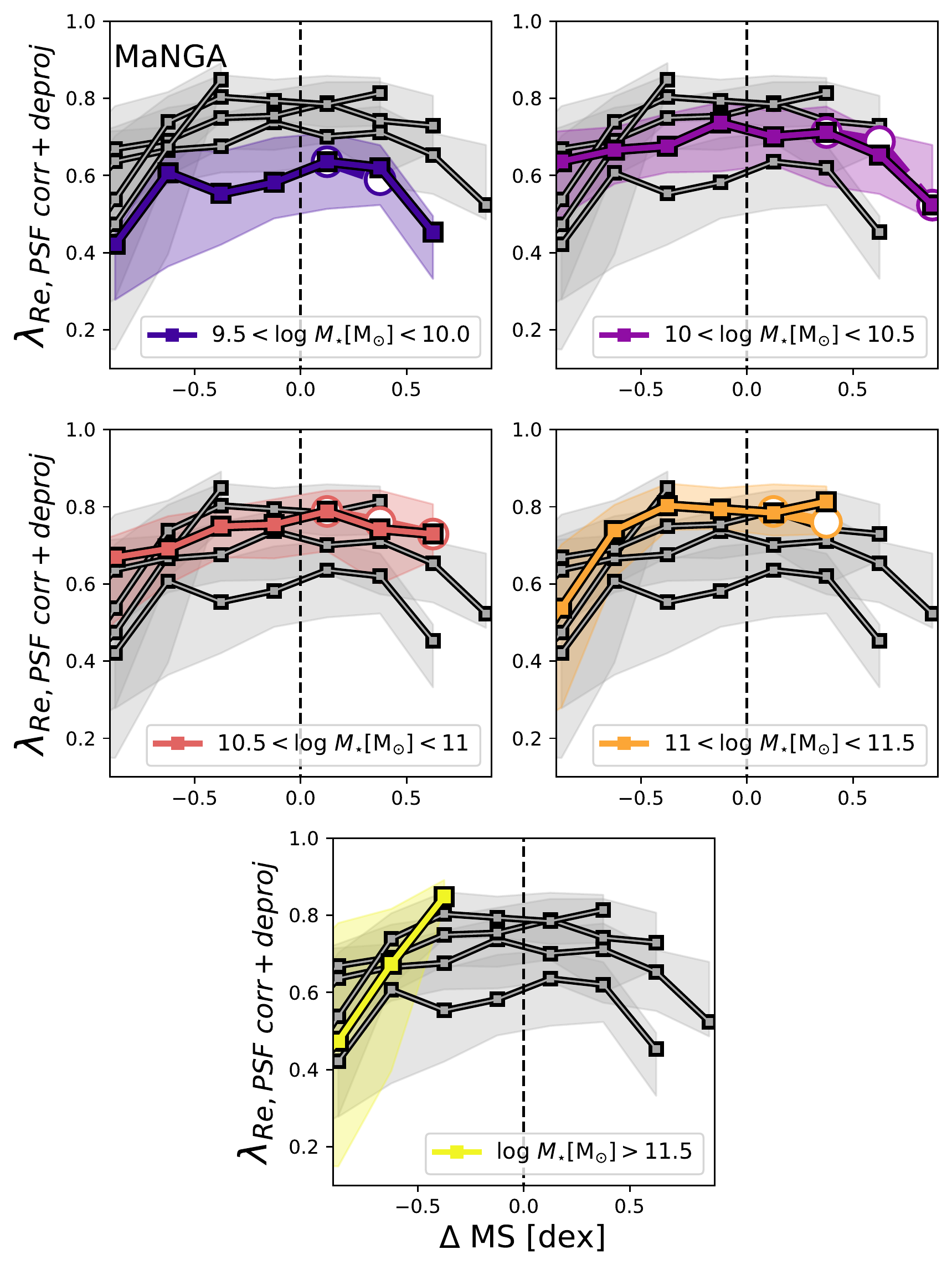}  
\end{subfigure}
\caption{Same as Figure~\ref{results2_MaNGA}, but using the linear main sequence line of Equation~\ref{eqn2} to calculate $\Delta~\rm{MS}$. Given there are fewer galaxies above the linear SFMS line, we are not able to probe as far above the main sequence as with the curved line.}
\label{A2}
\end{figure*}

We present median $\lambda_{Re}$ in bins of stellar mass as a function of $\Delta~\rm{MS}$ using the \textit{linear} definition of the SFMS line from Equation~\ref{eqn2}. Figure~\ref{A1} shows the SAMI results, and~\ref{A2} are the MaNGA results. 
 
 It is worth noting that the increase of $\lambda_{Re}$ with stellar mass at the locus of the main sequence described in Section~\ref{results} remains even if $\Delta~\rm{MS}$ is measured from the linear fit to the SFMS. The only difference is the change in behaviour at the highest stellar mass bins, simply because we no longer have galaxies at these stellar masses on the SFMS.
 
Interestingly, the trend of a steepening of the $\Delta~\rm{MS}$--$\lambda_{Re}$ relation below the main sequence almost entirely disappears (or is at least pushed towards higher distances from the main sequence) when a linear fit to the main sequence is used.  

    \bibliographystyle{mnras}
  \bibliography{kinbib.bib}
\end{document}